\documentclass[12pt]{article}

\usepackage{amsfonts}
\usepackage[english]{babel}
\usepackage{graphicx}
\usepackage{pstricks}
\usepackage{times}

\begin{document}

\begin{titlepage}
\begin{flushright}
TPI-MINN-00/53-T  \\
UMN-TH-1929/00   \\
FTUV/001113 \\
IFIC/00-70 \\
\end{flushright}

\vspace{0.6cm}

\begin{center}
\Large{{\bf Domain walls in supersymmetric QCD:}}\\
\Large{{\bf The taming of the zoo}}
\end{center}

\vspace{1cm}

\begin{center}
\large Daniele Binosi
\end{center}

\begin{center}
{\em
Departamento de F\'\i sica Te\'orica and IFIC, Centro Mixto, \\
Universidad de Valencia-CSIC,\\
E-46100, Burjassot, Valencia, Spain\\}
\end{center}

\vspace{0.1cm}

\begin{center}
\large Tonnis ter Veldhuis
\end{center}

\begin{center}
{\em Theoretical Physics Institute, Univ. of Minnesota, 
Minneapolis,
       MN 55455}
\end{center}
\vspace{0.5cm}

\begin{abstract}

We provide a unified picture of the domain wall spectrum in supersymmetric QCD
with  $N_c$ colors and $N_f$ flavors of quarks in the (anti-) fundamental 
representation. Within the framework of the  Veneziano-Yankielowicz-Taylor
effective Lagrangian, we consider domain walls connecting 
chiral symmetry breaking vacua, and we take the quark masses  to be 
degenerate. For $N_f/N_c<1/2$, there is one BPS saturated domain wall  for any
value of the quark mass $m$. For $1/2 \le N_f/N_c < 1$ there are two critical
masses, $m_*$ and $m_{**}$, which depend on the number of colors and flavors
only through the ratio $N_f/N_c$; if $m<m_*$, there are two BPS walls, if
$m_*<m<m_{**}$ there  is one non-BPS wall, and if $m>m_{**}$ there is no domain
wall.  We numerically determine $m_*$ and $m_{**}$ as a function of $N_f/N_c$,
and we find that $m_{**}$ approaches a constant value in the limit that this
ratio goes to one.

\end{abstract}

\vspace{0.5cm}

\begin{flushleft}
E-mail: binosi@hal.ific.uv.es; veldhuis@hep.umn.edu
\end{flushleft}
\end{titlepage}

\section*{Introduction}

In recent years, several groups have studied domain walls in supersymmetric
quantum chromodynamics (SQCD) in the framework of the 
Veneziano-Yankielowicz-Taylor (VYT) effective Lagrangian \cite{VY82TVY83}. This
Lagrangian is not an effective Lagrangian in the usual Wilsonian sense, but it
it is expected to describe the vacuum structure  of SQCD. If the masses of the
quark  matter fields are degenerate, the VYT effective Lagrangian takes the
form of a Wess-Zumino model in terms of the chiral superfields $\Phi$,
representing the glueball superfield, and $X$, representing the quark matter.
The effective Lagrangian has $N_c$ chiral symmetry breaking vacua and an
additional chiral symmetry conserving vacuum. The spectrum of the domain walls
connecting two adjacent chiral symmetry breaking vacua was found to depend on
the number of colors $N_c$, the number of flavors $N_f$, and the common mass of
the quarks $m$. In Ref.\cite{Sm97}, domain walls were studied in several models
with $N_f=N_c-1$. In the cases of $N_c=2,3$ and $4$, it was found that for 
small quark mass $m$, there are two Bogomolnyi-Prasad-Sommerfield (BPS)
saturated domain walls interpolating between two adjacent vacua. Above a first
critical mass $m_*$, these BPS walls combine and form one non-BPS wall. If the
mass is higher than a second critical value $m_{**}$, there is no domain wall
left; the two adjacent vacua can only be connected by a two wall configuration
passing through the chiral symmetry conserving vacuum. In Ref.\cite{dCM99} it
was found that for any theory with $N_f/N_c<1/2$ there is one BPS wall
connecting adjacent chiral symmetry breaking vacua both in the small $m$ limit
and  in the large $m$ limit.

The purpose of this work is to provide a unified description of the domain wall
spectrum of the VYT effective Lagrangian  for any value of $N_c$ and $N_f$. One
key observation we make is that, with a canonical K\"ahler potential, the
static solutions to the equations of motion take the form
\begin{eqnarray}
\phi & = & \phi(N_c z, m, N_f/N_c), \nonumber\\
\chi & = & \chi(N_c z, m, N_f/N_c).
\label{solutionform}
\end{eqnarray}
Here $\phi$ and $\chi$ are the scalar components of the superfields $\Phi$ and
$X$ respectively. Theories with different values of $N_c$ but identical ratios
of $N_f/N_c$ have domain wall solutions that differ only by a scale factor
$N_c$. Therefore, only theories with mutually prime values of $N_c$ and $N_f$
need to be considered. Theories for which $N_c$ and $N_f$ have common divisors
are related to theories with lower values of $N_c$ and $N_f$.

We also make use of the fact that BPS states lie in shortened multiplets of the
centrally extended supersymmetry algebra. Only two shortened multiplets can
combine to form one regular multiplet, and therefore the number of BPS
saturated domain walls can only change by  a multiple of two  when the mass
parameter is continuously varied \cite{Ritz:2000xa}. For small values of $m$,
we show that if $N_f/N_c<1/2$ there is one BPS wall, and if \mbox{$1/2 \le
N_f/N_c <1$} there are two BPS walls. It then follows that the number of BPS
walls is odd for any value of $m$ if $N_f/N_c<1/2$, and even for any value of
$m$ if $1/2 \le N_f/N_c < 1$. Note that this is fully consistent with the
previous  results in Refs.\cite{Sm97,dCM99}.

Even though the single BPS wall at small $m$ for $N_f/N_c<1/2$ can not
disappear when the mass is increased, one can not exclude the possibility of
the appearance and subsequent disappearance of a new pair of BPS walls at
intermediate values of $m$. We therefore perform a numerical study of this
class of theories, and find that such an exotic event does not occur.

For theories with $1/2 \le N_f/N_c <1$, the fact that the number of BPS wall is
even does not block the disappearance of the pair of degenerate BPS walls that
exist at small $m$.  Indeed, in the specific cases previously studied 
\cite{Sm97} this happens at the mass $m_*$.  We also perform a numerical study
of the theories in this class, and find that the prototype behavior found in
Ref.\cite{Sm97} is valid in general.  These theories all have a critical mass
$m_*$ at which the two BPS walls combine to form a non-BPS wall, and a second
critical mass $m_{**}$, where the non-BPS wall disappears. It is clear from the
general form of the static solutions in  Eq.(\ref{solutionform}) that the
values of $m_*$ and $m_{**}$ only depend on $N_f$ and $N_c$ through the ratio
$N_f/N_c$. 

The remainder of this paper is organized as follows. In Section \ref{SQCDVYT}
we introduce the VYT Lagrangian as an effective description of SQCD. Next, in 
Section \ref{walls} we discuss domain walls in the VYT model, followed by the
presentation of our numerical results in  Section \ref{lesshalf} and
\ref{morehalf}.  We first show that in the theories with $N_f/N_c<1/2$ there is
one BPS wall for all values of $m$. Then, for the theories with $1/2 \le
N_f/N_c <1$, we determine  $m_*$ and $m_{**}$ as a function of $N_f/N_c$. In
the final section we discuss how our results depend on some of the assumptions
we made, and to  what extend the results we derived in the context of the VYT
effective Lagrangian apply to SQCD.

\section{Supersymmetric QCD and the VYT effective Lagrangian
\label{SQCDVYT}}

The fundamental Lagrangian describing SQCD, extensively discussed
in for example the review \cite{Shifman:1997ua}, takes the form
\begin{eqnarray}
{\cal L} & = & \left(\frac 1{4g^2}\,{\mathrm{tr}}\int\!d^2\theta\,W^2+
{\mathrm{h.c.}}\right)
+\left(\frac14
\int\!d^4\theta\, Q^{i\dagger}\,{\mathrm e}^V Q_i+\frac14
\int\!d^4\theta\,\bar Q^{j\dagger}\,{\mathrm e}^{-V} \bar Q_j
\right) \nonumber \\
& & -\left(\frac{m^{ij}}2\int\!d^2\theta\, \bar Q_i Q_j
+{\mathrm{h.c.}}\right).
\end{eqnarray}
Here color indices have been suppressed, and $i=1,\dots,N_f$ is a flavor
index.  The vector superfield $V$ contains the gauge bosons and gauginos, and
$Q_i$ and $\bar Q_i$ are the quark chiral  superfields transforming
respectively under the $N_c$ and $\bar N_c$ representations of the $SU(N_c)$
gauge  group. In addition, $g$ is the gauge coupling, and $m^{ij}$ is the quark
mass matrix. If $N_f \le N_c-1$ then the theory has a dynamically generated
superpotential. In the absence of a quark mass, the model has a $SU(N_f)\otimes
SU(N_f) \otimes U(1)_V \otimes U(1)_A \otimes U(1)_R$ global symmetry at the
classical level. The $U(1)_R$ symmetry is broken down to the discrete symmetry
${\mathbb Z}_{2N}$ by an anomaly. It is further broken down spontaneously to
${\mathbb Z}_2$ when a gaugino condensate  $({\mathrm{tr}}\,\lambda\lambda)$ is
formed.

The VYT effective Lagrangian proposed in Ref.\cite{VY82TVY83} and modified in
Ref.\cite{KS97} to properly realize the discrete anomaly free  ${\mathbb
Z}_{2N}$  symmetry, provides a  partial  description of SQCD with $N_f<N_c$ at
low energy in terms of the  composite chiral superfields
\begin{equation} 
\Phi^3=\frac3{32\pi^2}\,{\mathrm{tr}}\,W^2, \qquad
M_{ij}=2\bar Q_i Q_j,
\end{equation} 
representing the glueball and matter chiral superfields, respectively.

The superpotential of the model, which is constructed to
reproduce the anomalous Ward identities of SQCD, takes the form
\begin{equation} {\cal
W}=\frac23\Phi^3\left[\log\left(\frac{\Phi^{3(N_c-N_f)}{\mathrm{det}}M}
{\Lambda^{3N_c-N_f}}\right)-(N_c-N_f)\right]
-\frac12\,{\mathrm{tr}}\left(mM\right).
\end{equation} 
The VYT effective Lagrangian is not an effective theory in the usual Wilsonian
sense. It does not describe all relevant low energy degrees of freedom, but it
reproduces Green's functions involving only the $\Phi$ and the $M$ fields
\cite{KS97}.

We take the quark mass matrix to be diagonal and degenerate, $m^{ij}=m
\delta^{ij}$. We then assume that for the purpose of studying the vacuum
structure and domain walls, we can take
$M_{ij}=X^2\delta_{ij}$. In terms of X, the superpotential reads
\begin{equation} 
{\cal
W}=\frac23\Phi^3\left[\log\left(\frac{\Phi^{3(N_c-N_f)}X^{2N_f}}
{\Lambda^{3N_c-N_f}}\right)-(N_c-N_f)\right]-\frac
m2N_fX^2. \label{VYTsuperpotential}
\end{equation}
The K\"ahler potential is not uniquely determined  by symmetry considerations. 
For now, we take it to have the canonical form
\begin{equation}
{\cal K}=\bar\Phi\Phi + \bar XX.
\label{kahler}
\end{equation} 
In Section \ref{discussion} we will comment on how our results may change with
a different form of the K\"{a}hler potential. The kinetic terms for the scalar
components are
\begin{equation} 
{\cal
L}_{\mathrm{kin}}=\partial^\mu\phi\partial_\mu\phi^*+\partial^\mu\chi 
\partial_\mu\chi^*,
\end{equation}
and the scalar potential is ($\Lambda=1$ from here on) 
\begin{equation} 
V\left(\phi,\chi\right)=
4\left|\phi^2\log\left[\phi^{3(N_c-N_f)}\chi^{2N_f}\right]\right|^2
+N_f^2\left|m\chi-\frac{4\phi^3}{3\chi}\right|^2.
\label{scalarpot}
\end{equation}
This potential has $N_c$ chiral symmetry breaking vacua, corresponding to
the physical expectation values
\begin{eqnarray}
\langle\phi^3\rangle_l & = &\left(\frac{3m}4\right)^{N_f/N_c}
{\mathrm e}^{2\pi il/N_c}, 
\nonumber \\ 
\langle\chi^2\rangle_l & = & \left(\frac4{3m}\right)^{(N_c-N_f)/N_c}
{\mathrm e}^{2 \pi i l /N_c}, \label{physvacuum}
\end{eqnarray} 
where $l=0,..,N_c-1$. In this paper we consider domain walls that interpolate
between such vacua. In order to ensure that the two connected vacua lie on the
same sheet of the logarithm in the scalar potential (\ref{scalarpot}), we write
the vacuum expectation values of $\phi$ and $\chi$ as
\begin{eqnarray} 
\langle\phi\rangle_k & = &
\langle\phi\rangle_0{\mathrm e}^{-2\pi i
kN_f/3N_c}, 
\qquad 
\langle\phi\rangle_0\equiv\left(\frac{3m}4\right)^{N_f/3N_c}, \nonumber \\ 
\langle\chi\rangle_k & = & \langle\chi\rangle_0{\mathrm
e}^{\pi ik(N_c-N_f)/N_c}, \qquad 
\langle\chi\rangle_0\equiv\left(\frac4{3m}\right)^{(N_c-N_f)/2N_c}, 
\label{minchi} 
\end{eqnarray} 
with $k=0,..,N_c-1$. This ensures that the phase  of the argument of the 
logarithm is equal to zero in the vacuum. Note that not all vacua  can be
written in the form of Eq.(\ref{minchi}). For example, the vacuum with $l=1$ in
Eq.(\ref{physvacuum})  in the case $N_c=4$ and $N_f=2$ does not lie on the same
sheet as the vacuum with $l=0$. The latter vacuum is obtained both for $k=0$ or
$2$ in Eq.(\ref{minchi}). These two vacua can therefore not be connected by a
domain wall, unless it crosses the branch cut. The SQCD gaugino condensates in
the vacuum labeled by $k$ is
\begin{equation}
\langle{\mathrm{tr}}\,\lambda\lambda\rangle_k=
\frac{32 \pi^2}{3} \langle \phi^3\rangle_k=
\frac{32 \pi^2}{3} \left(\frac{3m}{4}\right)^{N_f/N_c}
{\mathrm e}^{- 2\pi i k N_f/N_c}.
\label{condensate}
\end{equation}

Apart from the chiral symmetry breaking vacua,  the VYT model also has a
symmetric vacuum  with $\phi=0$ and $\chi=0$, corresponding to vanishing
gaugino condensate \cite{KS97}.  The existence of a corresponding vacuum in
SQCD is very controversial.

In terms of the physical field $S=\Phi^3$, it is clear that the chirally
symmetric vacuum in the VYT Lagrangian arises from  a singularity in the metric
that is derived from the K\"{a}hler potential (\ref{kahler}), whereas the
chiral symmetry breaking vacua are associated with extrema of the
superpotential (\ref{VYTsuperpotential}).  As noted before, the K\"{a}hler
potential is not completely determined by symmetry considerations, in contrast
to the superpotential. Results that depend on the particular details of the 
K\"{a}hler potential are therefore suspect.

In favor of the chirally symmetric vacuum, it was argued in Ref.\cite{KS97}
that the existence of such a vacuum in combination with a vacuum averaging
hypothesis could resolve a long standing inconsistency concerning gluino
condensates calculated using the strong coupling instanton approach on the one
side, and the weak coupling instanton approach on the other. However, in
Ref.\cite{Ritz:2000mq} it was shown that a similar inconsistency persists in
supersymmetric QCD with light matter in the adjoint representation,  even with
the vacuum averaging hypothesis. As this theory is equivalent to softly broken
${\cal N}=2$ supersymmetric Yang-Mills theory (SYM), there is no uncertainty
about its vacua if the mass of the adjoint is small. The  result therefore
casts doubt on the motivation for the chiral symmetry breaking vacuum in
supersymmetric QCD.

In Ref.\cite{MS} it was stated that in the chiral symmetry conserving vacuum of
the VYT Lagrangian certain discrete anomaly matching condition can not be
satisfied. In Ref.\cite{Kogan:1998dt} on the contrary it was argued that  such
considerations do not constrain the existence of a symmetric vacuum.

Another piece of evidence pointing against the existence of a chirally
symmetric vacuum in SQCD stems from deformed ${\cal N}=2$ supersymmetric
theories.  ${\cal N}=2$ SYM  theory with a small mass for the adjoint has only
two vacua. ${\cal N}=1$ SYM is obtained from this theory in the  limit that the
soft breaking mass goes to infinity. In
Ref.\cite{Kaplunovsky:1999vt,Artstein:2000ke} it was thus argued that ${\cal
N}=1$ SYM also has only two vacua, barring a phase transition. This argument
indicates that there is no room for a third, chirally symmetric vacuum. A
similar argument concerning SQCD with one flavor can be made by deforming 
${\cal N}=2$ supersymmetric QCD with one hyper multiplet  \cite{Gorsky:2000ej}.

We conclude that it is dangerous to extend results obtained from the VYT
effective Lagrangian to SQCD if these results depend on the existence of the
chirally symmetric vacuum. In this article we only  consider domain walls
interpolating between chiral symmetry breaking vacua.

\section{Domain walls in the VYT effective Lagrangian
\label{walls}}

Domain walls are static field configurations that interpolate between two
vacua. They are static solutions to the Euler-Lagrange equations
\begin{equation}
\frac{\partial^2\phi}{\partial t^2}-
\frac{\partial^2\phi}{\partial z^2}=-\frac{\partial V}{\partial\phi^*}, \qquad
\frac{\partial^2\chi}{\partial t^2}-
\frac{\partial^2\chi}{\partial z^2}=-\frac{\partial V}{\partial\chi^*}.
\label{fullsystem}
\end{equation}
It can be checked that static solutions to these equations take the form of
Eq.(\ref{solutionform}). We will refer to the theory with $N_c$ colors and
$N_f$ flavors as $(N_c,N_f)$. It is clear that theories with the same ratio of
$N_f/N_c$ have domain walls that only differ by a scale factor. 

BPS saturated domain walls~\cite{Chibisov:1997rc} in the VYT model satisfy  the
first order equations
\begin{equation}
\frac{d\phi}{dz} = {\rm e}^{i\delta}\frac{\partial \bar{\cal W}}
{\partial\phi^*}, \qquad
\frac{d\chi}{dz} = {\rm e}^{i\delta}\frac{\partial \bar{\cal W}}
{\partial\chi^*}. 
\label{BPSsystem}
\end{equation}
Solutions to the BPS equation have an ``integral of motion''
\begin{equation}
I={\mathrm{Im}}\{{\mathrm e}^{-i\delta}{\cal W}\},
\label{iom}
\end{equation}
which does not depend on the space coordinate $z$.

We consider domain walls interpolating between adjacent vacua. For definiteness
we take the vacua for $k=0$ and $k=1$. Note that these vacua are only
neighboring in terms of the phase of the gaugino condensate if $N_f=1$ or
$N_f=N_c-1$. Evaluation of the constant of motion $I$ in these two vacua
provides the  value of the phase  $\delta$ in the BPS equation,
\begin{equation}
\delta=\pm\frac{\pi}2-\pi \frac{N_f}{N_c},
\end{equation}
where the choice of the sign determines which of the two vacua is located at
$z=+ \infty$, and which at $z=- \infty$. The BPS bound on the energy of the
interpolating domain wall is given by
\begin{equation}
E_{\mathrm{BPS}} = 2 N_c \left(\frac{4}{3}\right)^{1-N_f/N_c} m^{N_f/N_c} \sin
\left(\pi\frac{N_f}{N_c}\right).
\end{equation}
We parametrize the domain wall profiles in terms of four real functions by
means of
\begin{eqnarray}
\phi(z)&=& \langle\phi\rangle_0R(z){\mathrm e}^{i\beta(z)}, \\
\chi(z)&=& \langle\chi\rangle_0\rho(z){\mathrm e}^{i\alpha(z)}.
\end{eqnarray}
If the domain wall profiles have reflection symmetry with respect to their
center, then the integral of motion (\ref{iom}) evaluated at the center 
($z=0$) yields the following relation between the values of $R_0\equiv R(0)$
and $\rho_0\equiv\rho(0)$ 
\begin{equation}
-R_0^3\left[\left(1-\frac{N_f}{N_c}\right)\left(\ln R_0^3-1\right)+
\frac{N_f}{N_c}\ln\rho_0^2\right]-\frac{N_f}{N_c}\rho_0^2=
\cos\left(\pi\frac{N_f}{N_c}\right).
\label{0condition}
\end{equation}
Note that $N_c$ and $N_f$ enter in Eq.(\ref{0condition}) only through the ratio
$N_f/N_c$, and there is no dependence on $m$. A BPS saturated domain wall  in
the theory $(N_c,N_f)$ correspond to a point on a curve defined by
Eq.(\ref{0condition}) in the $\rho_0^2$ versus $R_0^3$ plane. There is one such
curve for every value of $N_f/N_c$. In order to determine exactly where on the
curve the point is located, the BPS equation needs to be solved explicitly.
When the mass $m$ is varied and the domain wall remains BPS saturated, the
point representing the wall will move along its curve. 

\setlength{\unitlength}{.5mm}
\begin{figure}[!t]
\begin{center}
\includegraphics[width=8.0cm]{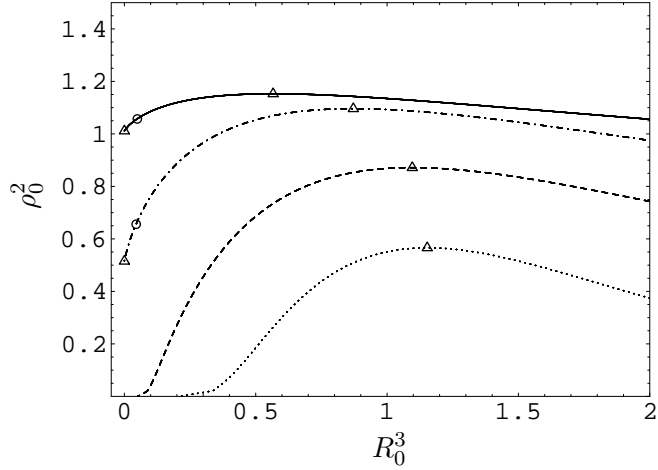}
\put(-172,60){\small{\rotateleft{$\rho_0^2$}}}
\put(-76,-6){\small{$R_0^3$}}
\caption{The solution to Eq.(\ref{0condition}) for the {\it (5,1)} (dotted),
{\it (5,2)} (dashed), {\it (5,3)} (dotted-dashed) and {\it (5,4)} (continuous)
theory. Open triangles indicate the \mbox{$m=0$} domain wall solution(s), while
open circles represent the domain wall solution corresponding to
\mbox{$m=m_*$}.}
\label{fig1}
\end{center}
\end{figure}

We now consider the small mass limit $m\ll 1$, as previously discussed in
Refs.\cite{Kovner:1997ca,dCM99}. In this limit the field $\Phi$ can be
integrated out by imposing the condition $\partial{\cal W}/\partial\phi=0$. In
the resulting theory, which only  contains the field $X$, there is one BPS wall
interpolating between the $k=0$ and $k=1$ vacua.  This wall corresponds  to the
maximum\footnote{It is amusing to observe that the maxima of the curves of the
$(N_c,N_f)$ and $(N_c,N_c-N_f)$ theories are related by reflection in the line
$\rho_0^2=R_0^3$.} of the curve $\rho_0^2$ versus $R_0^3$ defined by
Eq.(\ref{0condition}).  However, for certain values of $N_c$ and $N_f$ there is
a second possible BPS wall in the small mass limit for which $\phi\equiv0$.
This wall corresponds to the point
\begin{equation}
R^3_0=0, \qquad 
\rho^2_0 =-\frac{N_c}{N_f}\cos\left(\pi\frac{N_f}{N_c}\right).
\end{equation}
As $\rho_0$ is a modulus and therefore larger than or equal to zero, the second
BPS wall is only present when $N_f/N_c \ge 1/2$.

We thus see that for small $m$ there is one BPS saturated wall if
$N_f/N_c<1/2$, and there are two BPS walls if $1/2 \le N_f/N_c <1$. Therefore, 
if  $N_f/N_c<1/2$ the number of BPS walls must be odd for {\it any} value of
the mass parameter $m$. In particular, the wall at small $m$ can not just
disappear. Similarly, if $1/2\le N_f/N_c<1$, the number of BPS walls must be
even.  This allows the possibility of the simultaneous disappearance of the
pair of BPS domain walls that exists at small $m$.

We illustrate this analysis in Fig.~\ref{fig1}  for the theories with $N_c=5$
and $N_f=1,2,3$ and $4$.  We draw the four curves given by
Eq.(\ref{0condition}) in the $\rho_0^2$ versus $R_0^3$ plane, and indicate the
location of the BPS walls at small $m$ on the curves with a triangle.  For
$N_f=1$ and $2$, there is only one triangle on the curve. When the mass $m$ is
increased, these triangles start to move along the curve. How they move is a
dynamical question which can only be answered by explicitly solving the BPS
equations. It is clear though that they have to remain on their curve, they
have nowhere else to go. In principle, there is no {\it a priori} argument that
precludes the  appearance of a new pair of BPS walls. As we will show in
Section  \ref{lesshalf}, this does not occur.  For $N_f=3$ and $4$ there are
two triangles on the curves. As we will show in Section \ref{morehalf}, when
$m$ is increased, the two triangles start to move towards each other, and at
the  critical value $m_*$ they meet at the point on the curve indicated by a
circle. They then combine to form a non-BPS wall.

\section{The case $N_f/N_c<1/2$ \label{lesshalf}}

The existence of one BPS wall at small $m$ for theories with $N_f/N_c<1/2$, in
combination with the fact that the number of BPS walls modulo two is conserved,
implies that the number of BPS walls in this class of theories is odd for any
value of $m$. In order to determine whether any new pairs of BPS walls or
additional non-BPS walls appear  when $m$ is increased from zero, we performed
a numerical simulation of all the theories in this class with $N_c\le 8$.

We simulated the second order equations of motion on a lattice with a forward
predicting algorithm. To that purpose, we transformed  the Euler-Lagrange
differential Eqs.(\ref{fullsystem}) to finite difference equations, with
discretized space and time coordinates.  The scalar field $\phi$ and $\chi$
were kept fixed at their  vacuum expectation values at the two ends of the
lattice. The size of the lattice is chosen to be much larger than the width of
the walls and, at the  same time, the lattice spacing is chosen much smaller
than the width. An initial configuration for $\phi$ and $\chi$ was constructed
so that it interpolates between the two vacuum  expectation values. This
initial configuration does not satisfy the static equations of motion and
therefore evolves in time. We are interested in a minimum energy configuration.
The system can only reach such a minimum energy configuration from the initial
configuration if it can dissipate energy. We therefore added a friction term to
the equations of motion. The coefficient of friction is adjusted so that the
system relaxes to the minimum energy configuration in as short a time as
possible. If the coefficient is too large, the system does not converge, and if
it is too small, the systems keeps oscillating. Of course, as the friction
terms are proportional to the time derivatives of the fields, the static
configurations of the equations of motion with and without the friction term
are the same. Once the system had come to rest, we calculated the energy of 
the final  configuration and compared this to the BPS bound. We also explicitly
checked  that domain wall solutions we thus  obtained did not cross the  branch
cut of the logarithm appearing in the scalar potential (\ref{scalarpot}). This
check was performed by plotting the  argument of the logarithm,
$\phi^{3(N_c-N_f)}\chi^{2N_f}$, in the complex plane. As the fields on the
domain wall vary from one vacuum to the other, this quantity traces a loop in
the complex plane. We then verified that this loop does not cross the negative
real axis (where the  branch cut is located within our conventions).

\setlength{\unitlength}{.5mm}
\begin{figure}[!t]
\begin{center}
\includegraphics[width=13.0cm]{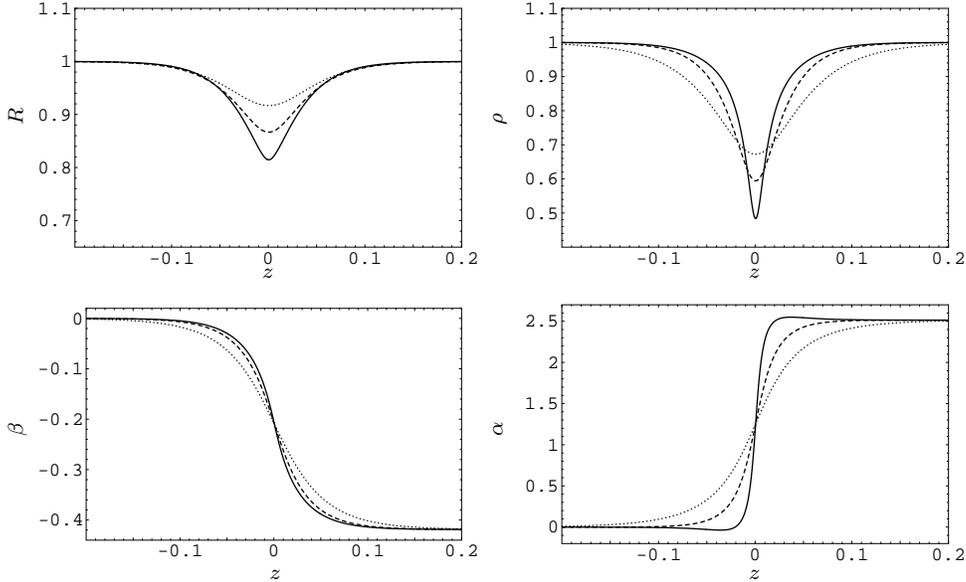}
\put(-261,123){\scriptsize{\rotateleft{$R$}}}
\put(-193.5,82){\scriptsize{$z$}}
\put(-132,123){\scriptsize{\rotateleft{$\rho$}}}
\put(-64,82){\scriptsize{$z$}}
\put(-261,40){\scriptsize{\rotateleft{$\beta$}}}
\put(-191.5,2){\scriptsize{$z$}}
\put(-132,40){\scriptsize{\rotateleft{$\alpha$}}}
\put(-64,2){\scriptsize{$z$}}
%
\caption{BPS domain walls for the {\it (5,1)} theory.
Masses are $m=10$ (dotted), 20 (dashed) and 50 (solid)}
\label{fig2}
\end{center}
\end{figure}

We repeated this procedure for each theory, starting at low values of $m$ and
following the walls as $m$ is increased. In order to see if more than one wall
exists, we also repeated the procedure with various initial conditions. We
found that for the theories in the class $N_f/N_c<1/2$ there is only one BPS
saturated domain wall for any value of $m$. In Fig.~\ref{fig2} we show some
representative examples of BPS domain walls for the {\it (5,1)} theory and
various values of $m$. Following the  domain wall in this theory in
Fig.~\ref{fig1}, it starts on top of its curve at $m=0$, and then moves along
the curve towards smaller values of $R_0^3$ when $m$ is increased.

\section{The case $1/2\le N_f/N_c < 1$ \label{morehalf}}

If $1/2\le N_f/N_c<1$ there exist two BPS domain walls in the small $m$ limit.
The argument that the number of BPS walls can only change by a multiple of two
allows for the possible existence of two critical masses, $m_*$ and $m_{**}$,
as found in Ref.\cite{Sm97} for the theories $(N_c,N_c-1)$ and $N_c=2,3$ and
$4$. At the mass $m_*$, the two BPS walls combine to form one non-BPS wall, and
at the mass $m_{**}$ the non-BPS wall disappears. We perform a numerical
simulation of all theories in the class $1/2 \le N_f/N_c<1$ with $N_c\le 8$,
and find the same sequence in each theory. The theories we simulated provide a
sample of values of $N_f/N_c$ that covers the whole allowed range of $1/2\le
N_f/N_c < 1$. We therefore conclude that the theories in this class, not just
the ones with $N_c\le8$, behave qualitatively in the same way. Of course, the
actual values of $m_*$ and $m_{**}$ depend on $N_f/N_c$.

\setlength{\unitlength}{.5mm}
\begin{figure}[!t]
\begin{center}
\includegraphics[width=8.0cm]{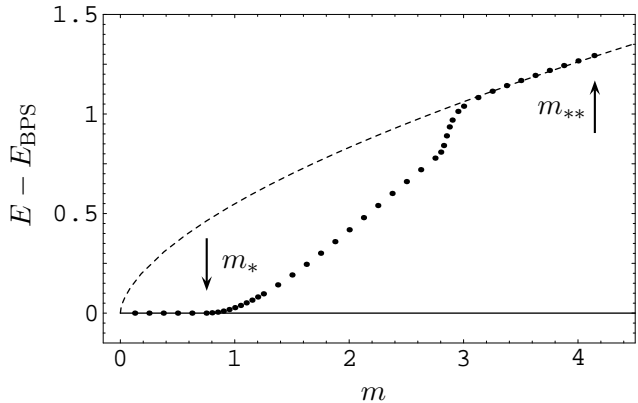}
\put(-168,39){\small{\rotateleft{$E-E_{\mathrm{BPS}}$}}}
\put(-75,-7){\small{$m$}}
\psline[arrows=->,arrowscale=0.9](-0.64,3.2)(-0.64,3.9)
\psline[arrows=->,arrowscale=0.9](-5.8,1.8)(-5.8,1.1)
\put(-112,28){\small{$m_*$}}
\put(-28,69){\small{$m_{**}$}}
\caption{The difference between the energy $E$ of the domain wall configuration
and the BPS bound $E_{\rm BPS}$ as a function of the mass $m$ (dots) for the
{\it (5,3)} theory. The dashed line indicates the difference $E_{2\rm W}-E_{\rm
BPS}$.}
\label{newfig}
\end{center}
\end{figure}

In Fig.~\ref{newfig} we show how the difference between the energy of the
domain wall and the BPS bound depends on the mass $m$ in a representative case,
the {\it(5,3)} theory. We have indicated the masses $m_*$ and $m_{**}$, and we
also plot the sum of the energy of the two BPS walls that connect the two vacua
for $k=0$ and $k=1$ through the intermediate vacuum at  $\phi=0$ and $\chi=0$,
\begin{equation} E_{2\mathrm W} = 2 N_c \left(\frac{4}{3}\right)^{1-N_f/N_c}
m^{N_f/N_c}. \end{equation} Below $m_*$, the energy of each of the two possible
walls is equal to the BPS bound. When the mass is increased above $m_*$, the
energy of the non-BPS wall starts to increase and tends towards $E_{2\rm W}$.
In fact, at a mass below $m_{**}$ the energy of the wall exceeds $E_{2\rm W}$.
However, at that point there is still a sphaleron barrier, and the wall is
meta-stable. The height of the barrier decreases as the mass approaches
$m_{**}$, and when the mass is equal to $m_{**}$, the non-BPS wall ceases to
exist.

Let us pause briefly to account for the kink in between $m_*$ and $m_{**}$
appearing in Fig.~\ref{newfig}. At this value of the mass parameter $m$ the
wall changes rapidly from a shape resembling the BPS configuration, to a shape
more similar to a two wall configuration approaching the vacuum in the middle.

In order to find $m_*$ numerically, we start with a small value of the mass
parameter $m$ for which the energy of the domain wall configuration does not
differ significantly from the BPS bound.  The initial configuration determines
which of the two possible walls we obtain; this does not matter.  Then we
gradually increase $m$ and determine $m_*$ as the value of $m$ for which the
energy of the wall differs significantly from the BPS bound. 

\setlength{\unitlength}{.5mm}
\begin{figure}[!t]
\begin{center}
\includegraphics[width=8.0cm]{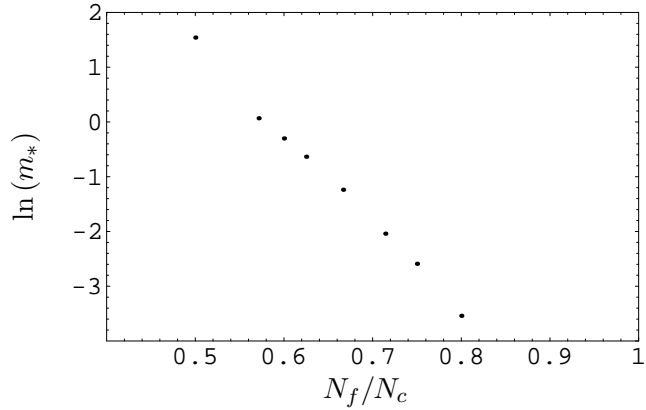}
\put(-172,40){\small{\rotateleft{$\ln\left(m_{*}\right)$}}}
\put(-89,-7){\small{$N_f/N_c$}}
\caption{Logarithm of the critical mass $m_*$ versus the
ratio $N_f/N_c$.}
\label{fig3}
\end{center}
\end{figure}

In Table~\ref{table1} we present the numerically determined values of the 
critical mass $m_*$ for the $(N_c,N_f)$ theories with $N_c\le8$. We also plot
the logarithm of $m_*$ as a function of the ratio $N_f/N_c$ in Fig.~\ref{fig3}.
We are not able to determine $m_*$ numerically for $N_f/N_c>4/5$ because its
value becomes too small. However, from Fig.~\ref{fig3} it appears that this
critical mass goes to zero in the limit $N_f/N_c\to1$.

In order to determine the critical mass $m_{**}$, we use the following
procedure. For a fixed value of $m$ we calculate the minimum energy $E$ for
field configurations interpolating between the given vacua subject to a
constrained value of $R_0$.  If $m$ is just above $m_*$, a plot of $E$ versus
$R_0$ shows a minimum and a maximum, corresponding to a non-BPS domain wall and
a sphaleron, respectively. As $m$ increases the minimum and the maximum move
towards each other and become less pronounced. We determine $m=m_{**}$ as the
value of $m$ were the maximum and the minimum coincide.  For higher values of
$m$, the energy $E$ increases monotonously with $R_0$, which means  that the
non-BPS wall has disappeared and the minimum energy configuration is formed by
two walls connecting through the chirally symmetric vacuum.

\begin{figure}[!t]
\begin{center}
\includegraphics[width=13cm]{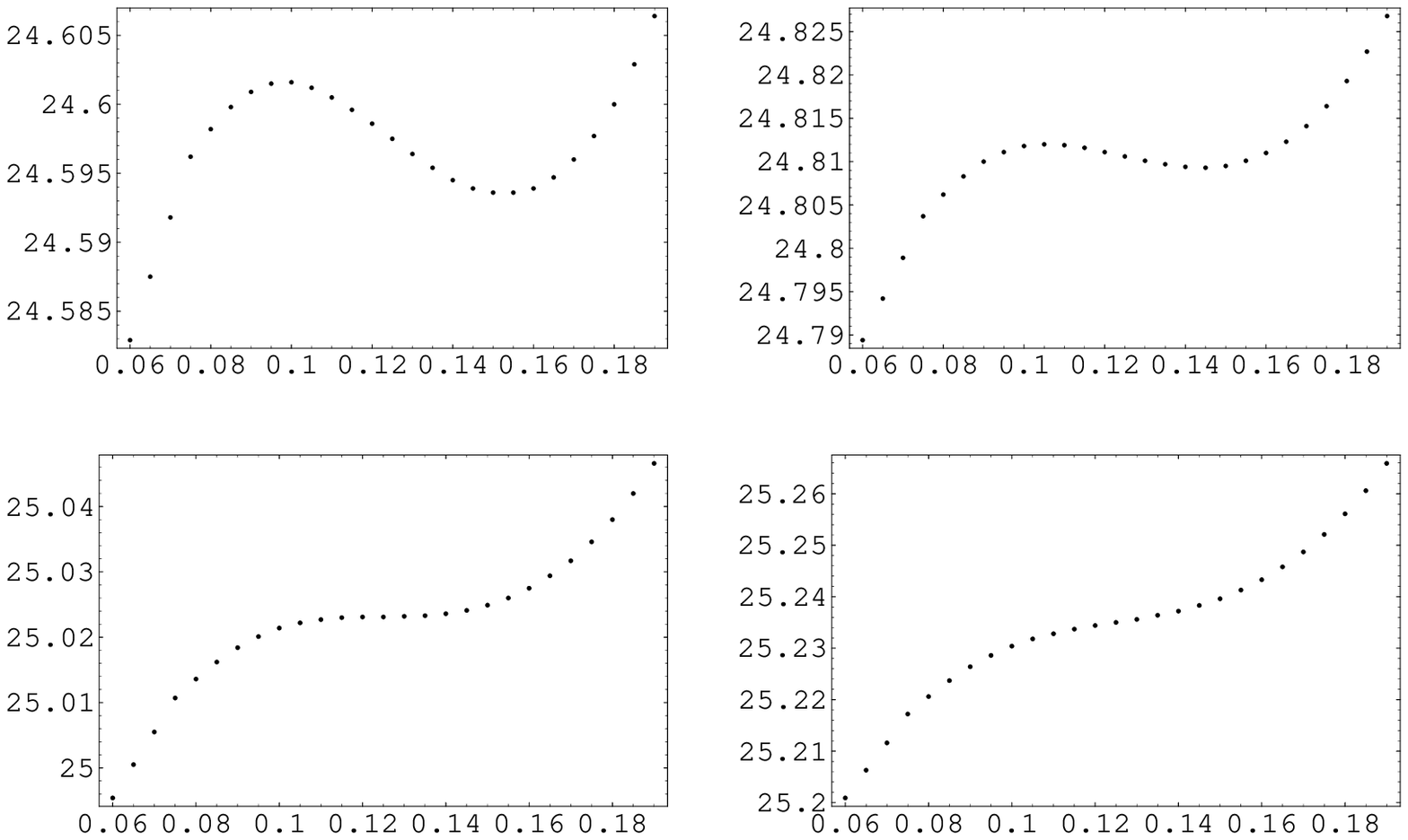}
\pspolygon[linecolor=white,fillstyle=solid,fillcolor=white]
(-6.2,6.2)(-5.35,6.2)(-5.35,6.5)(-6.2,6.5)
\pspolygon[linecolor=white,fillstyle=solid,fillcolor=white]
(-6.2,7.0)(-5.35,7.0)(-5.35,7.2)(-6.2,7.2)
\pspolygon[linecolor=white,fillstyle=solid,fillcolor=white]
(-6.2,5.4)(-5.35,5.4)(-5.35,5.7)(-6.2,5.7)
\pspolygon[linecolor=white,fillstyle=solid,fillcolor=white]
(-6.2,4.7)(-5.35,4.7)(-5.35,5.0)(-6.2,5.0)
\pspolygon[linecolor=white,fillstyle=solid,fillcolor=white]
(-6.2,0.4)(-6.2,0.8)(-5.5,0.8)(-5.5,0.4)
\pspolygon[linecolor=white,fillstyle=solid,fillcolor=white]
(-5.0,0.25)(-4.8,0.25)(-4.8,0.58)(-5.0,0.58)
\pspolygon[linecolor=white,fillstyle=solid,fillcolor=white]
(-4.2,0.25)(-3.95,0.25)(-3.95,0.58)(-4.2,0.58)
\pspolygon[linecolor=white,fillstyle=solid,fillcolor=white]
(-3.5,0.25)(-3.25,0.25)(-3.25,0.58)(-3.5,0.58)
\pspolygon[linecolor=white,fillstyle=solid,fillcolor=white]
(-2.8,0.25)(-2.55,0.25)(-2.55,0.58)(-2.8,0.58)
\pspolygon[linecolor=white,fillstyle=solid,fillcolor=white]
(-2.05,0.25)(-1.8,0.25)(-1.8,0.58)(-2.05,0.58)
\pspolygon[linecolor=white,fillstyle=solid,fillcolor=white]
(-1.3,0.25)(-1.05,0.25)(-1.05,0.58)(-1.3,0.58)
\pspolygon[linecolor=white,fillstyle=solid,fillcolor=white]
(-12.2,0.25)(-12.0,0.25)(-12.0,0.58)(-12.2,0.58)
\pspolygon[linecolor=white,fillstyle=solid,fillcolor=white]
(-11.4,0.25)(-11.2,0.25)(-11.2,0.58)(-11.4,0.58)
\pspolygon[linecolor=white,fillstyle=solid,fillcolor=white]
(-10.6,0.25)(-10.4,0.25)(-10.4,0.58)(-10.6,0.58)
\pspolygon[linecolor=white,fillstyle=solid,fillcolor=white]
(-10.0,0.25)(-9.8,0.25)(-9.8,0.58)(-10.0,0.58)
\pspolygon[linecolor=white,fillstyle=solid,fillcolor=white]
(-9.2,0.25)(-9.0,0.25)(-9.0,0.58)(-9.2,0.58)
\pspolygon[linecolor=white,fillstyle=solid,fillcolor=white]
(-8.5,0.25)(-8.3,0.25)(-8.3,0.58)(-8.5,0.58)
\pspolygon[linecolor=white,fillstyle=solid,fillcolor=white]
(-7.75,0.25)(-7.55,0.25)(-7.55,0.58)(-7.75,0.58)
\pspolygon[linecolor=white,fillstyle=solid,fillcolor=white]
(-12.0,4.4)(-11.85,4.4)(-11.85,4.63)(-12.0,4.63)
\pspolygon[linecolor=white,fillstyle=solid,fillcolor=white]
(-11.25,4.4)(-11.10,4.4)(-11.10,4.63)(-11.25,4.63)
\pspolygon[linecolor=white,fillstyle=solid,fillcolor=white]
(-10.5,4.4)(-10.35,4.4)(-10.35,4.63)(-10.5,4.63)
\pspolygon[linecolor=white,fillstyle=solid,fillcolor=white]
(-9.85,4.4)(-9.7,4.4)(-9.7,4.63)(-9.85,4.63)
\pspolygon[linecolor=white,fillstyle=solid,fillcolor=white]
(-9.15,4.4)(-8.95,4.4)(-8.95,4.63)(-9.15,4.63)
\pspolygon[linecolor=white,fillstyle=solid,fillcolor=white]
(-8.45,4.4)(-8.25,4.4)(-8.25,4.63)(-8.45,4.63)
\pspolygon[linecolor=white,fillstyle=solid,fillcolor=white]
(-7.7,4.4)(-7.55,4.4)(-7.55,4.63)(-7.7,4.63)
\pspolygon[linecolor=white,fillstyle=solid,fillcolor=white]
(-5.55,4.4)(-5.35,4.4)(-5.35,4.63)(-5.55,4.63)
\pspolygon[linecolor=white,fillstyle=solid,fillcolor=white]
(-4.8,4.4)(-4.63,4.4)(-4.63,4.63)(-4.8,4.63)
\pspolygon[linecolor=white,fillstyle=solid,fillcolor=white]
(-4.0,4.4)(-3.85,4.4)(-3.85,4.63)(-4.0,4.63)
\pspolygon[linecolor=white,fillstyle=solid,fillcolor=white]
(-3.4,4.4)(-3.25,4.4)(-3.25,4.63)(-3.4,4.63)
\pspolygon[linecolor=white,fillstyle=solid,fillcolor=white]
(-2.7,4.4)(-2.5,4.4)(-2.5,4.63)(-2.7,4.63)
\pspolygon[linecolor=white,fillstyle=solid,fillcolor=white]
(-2.0,4.4)(-1.8,4.4)(-1.8,4.63)(-2.0,4.63)
\pspolygon[linecolor=white,fillstyle=solid,fillcolor=white]
(-1.3,4.4)(-1.1,4.4)(-1.1,4.63)(-1.3,4.63)
\psline[arrows=->,arrowscale=0.9,linestyle=dashed,dash=3pt 
2pt](-10.25,6.1)(-10.25,6.8)
\psline[arrows=->,arrowscale=0.9](-8.4,7.0)(-8.4,6.3)
\psline[arrows=->,arrowscale=0.9,linestyle=dashed,dash=3pt
2pt](-3.6,5.5)(-3.6,6.2)
\psline[arrows=->,arrowscale=0.9](-2.2,7.2)(-2.2,6.5)
\psline[arrows=->,arrowscale=0.9,linestyle=dashed,dash=3pt 
2pt](-9.55,1.3)(-9.55,2.0)
\psline[arrows=->,arrowscale=0.9](-9.55,3.3)(-9.55,2.6)
\put(-172,100){\scriptsize{$m=2.86$}}
\put(-172,20){\scriptsize{$m=2.92$}}
\put(-40,100){\scriptsize{$m=2.89$}}
\put(-40,20){\scriptsize{$m=2.95$}}
\put(-190,82){\scriptsize{$R_0$}}
\put(-190,2){\scriptsize{$R_0$}}
\put(-60,82){\scriptsize{$R_0$}}
\put(-60,2){\scriptsize{$R_0$}}
\put(-264,116){\scriptsize{\rotateleft{Energy}}}
\put(-264,36){\scriptsize{\rotateleft{Energy}}}
\put(-134,116){\scriptsize{\rotateleft{Energy}}}
\put(-134,36){\scriptsize{\rotateleft{Energy}}}
%
%
\caption{The minimum energy $E$ of interpolating field configurations
subject to a constrained value of $R_0$ in the {\it (5,4)} theory.
The arrows indicate the position of the non-BPS domain wall
(solid) and of the sphaleron (dashed). As $m$ increases, the minimum and
the maximum move towards each other while the height of the barrier
decreases. At $m_{**}$ the wall and the sphaleron coincide and then
they disappear.}
\label{fig4}
\end{center}
\end{figure}

\begin{figure}[!t]
\begin{center}
\includegraphics[width=8.0cm]{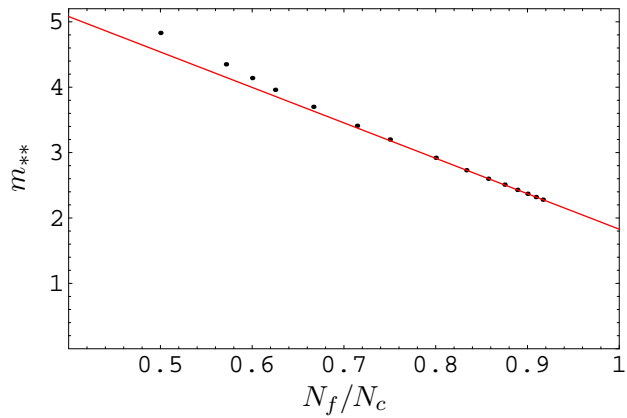}
\put(-166,50){\small{\rotateleft{$m_{**}$}}}
\put(-89,-7){\small{$N_f/N_c$}}
\caption{The critical mass $m_{**}$ versus the ratio
$N_f/N_c$. It can be seen that
$m_{**}$ approaches a constant value in the limit $N_f/N_c\to 1$. 
Extrapolating from our calculated points we find this limiting value
to be $m_{**}\approx 1.83$.}
\label{fig5}
\end{center}
\end{figure}

In Fig.~\ref{fig4} we illustrate this procedure for the {\it (5,4)} theory. The
numerically determined values of the critical mass $m_{**}$ for various
$(N_c,N_f)$ theories are presented in Table~\ref{table2}. In Fig.~\ref{fig5} we
plot $m_{**}$ versus the ratio $N_f/N_c$.  It can be seen from this figure that
in the limit $N_f/N_c \rightarrow 1$, $m_{**}$  tends to the constant value
$m_{**}\approx 1.83$.

\section{Discussion \label{discussion}}

We derived our results in the context of the VYT effective Lagrangian. To what
extend these results are valid for SQCD therefore depends on how well the VYT
effective Lagrangian describes the vacuum structure of SQCD.  An interesting
discussion of this issue is given in Ref.\cite{Kogan:1998dt}.

In addition, we assumed that the K\"{a}hler potential for the fields $\Phi$ 
and $X$ has the canonical form (\ref{kahler}). The tension of the BPS  walls
does not depend on the K\"{a}hler potential, but the shape of the domain walls
and quantities such  as the critical masses $m_*$ and $m_{**}$  do depend on
its details. It is instructive to see what happens in the case of a simple
deformation of the K\"{a}hler potential, 
\begin{equation} K = \alpha\left(\bar\Phi\Phi + \beta
\bar X X\right). 
\end{equation}  
After rescaling $z'=z/\alpha$, $X'=\sqrt{\beta}X$, $m'=m/\beta$ and
$\Lambda'=\Lambda \beta^{N_f/(3N_c-N_f)}$, the static equation of motions of
the scalar components are cast back into  exactly the same form they have when
$\alpha=\beta=1$, and our results apply to the primed quantities. Note that in
principle both $\alpha$ and $\beta$ can depend on $N_f$ and $N_c$, and not
necessarily only through the ratio $N_f/N_c$. Even when more complicated 
deformations of the K\"{a}hler potential are introduced, it is expected that
the qualitative picture remains intact, at least as long as the singularity
structure is not changed. In  particular, for any value of $m$, the number of
BPS walls will remain odd  for $N_f/N_c<1/2$, and even for $1/2 \le N_f/N_c <1$
under such deformations. Given the suspicion surrounding the  chirally
symmetric vacuum, it would be interesting to see  what happens if our analysis
were repeated with a K\"{a}hler potential that does not contain a singularity.
The fate of the lower branch for theories with $N_f/N_c>1/2$ is uncertain in
that case  \cite{Kaplunovsky:1999vt,Artstein:2000ke}, and our conclusions could
potentially be changed.

\setlength{\unitlength}{.5mm}

Finally, the difference in the domain wall spectrum of the theories with
$N_f/N_c <1/2$ and the theories with $1/2 \le N_f/N_c <1$ poses the following
question. What happens when the mass of one of the quark flavors in a theory
with $N_c=3$ and $N_f=2$ is moved to infinity? The theory with two degenerate
masses has an even number of BPS walls. If the mass of one of the quarks is
taken to infinity while the other mass is kept finite, the heavy quark can be
integrated out and the theory with $N_c=3$ and $N_f=1$ is obtained. However,
this latter theory has only one BPS wall. This change in the number of BPS
walls seems to  be inconsistent with the argument that the number of BPS walls
can only change by a multiple of two when a mass parameter is varied
continuously. Obviously,  one of the assumptions of the argument is violated.
We intend to come back to this issue in a future work. 

\section*{Acknowledgements}

The authors would like to thank A. Smilga, A. Vainshtein, B. de Carlos and V.
Vento for interesting discussions. T.t.V. would like to thank the Departamento
de F\'\i sica Te\'orica  of the Universidad de Valencia and the Instituto de
F\'\i sica Te\'orica  in Sao Paulo, where part of this work was done, for warm
hospitality. D.B. would like to thank the Theoretical Physics Institute of the
University of Minnesota, where the final part of this work was done, and
especially  Professor Mikhail Shifman, for warm hospitality and partial support
during his  visit. The research of T.t.V. is supported in part by the 
Department of Energy under Grant No. DE-FG-94ER40823. The work of D.B. is
supported by Ministerio de Educaci\'on y Cultura under Grant No.
DGICYT-PB97-1227.

\newpage

\begin{table}
\begin{center}
\begin{tabular}{c||c|c|c|c|c|c|c|}  
${\bf{N_f \Bigg{\backslash} N_c}}$ & {\it 2} & {\it 3} & {\it 4} & {\it 5} & 
{\it 6} & {\it 7} & {\it 8} \\
\hline \hline
{\it 1} & 4.67 & x & x & x & x & x & x \\ \hline
{\it 2} & & 0.29 & {\it (2,1)} & x & x & x & x \\ \hline
{\it 3} & & & 0.075 & 0.74 & {\it (2,1)} & x & x \\ \hline
{\it 4} & & & & 0.029 & {\it (3,2)} & 1.07 & {\it (2,1)} \\ \hline
{\it 5} & & & & & ? & 0.13 & 0.47 \\ \hline
{\it 6} & & & & & & ? & {\it (4,3)} \\ \hline
{\it 7} & & & & & & & ? \\ \hline
\end{tabular}
\caption{Numerically determined values of the critical mass $m_*$ for
different $(N_c,N_f)$ theories. The symbol `x' indicates that the 
corresponding theory has one BPS wall for any value of the mass parameter $m$
and therefore there is no $m_*$. The question mark indicates that the
value of $m_*$ is too small to be determined by our method.}
\label{table1}
\end{center}
\end{table}

\begin{table}
\begin{center}
\begin{tabular}{c|c|c|c|c|c|c|c|}  
${\bf{N_f\Bigg{\backslash}N_c}}$ & {\it 2} & {\it 3} & {\it 4} & {\it 5} & {\it 6} 
& {\it 7} & {\it 8} \\
\hline \hline
{\it 1} & 4.83 & x & x & x & x & x & x \\ \hline
{\it 2} & & 3.70 & {\it (2,1)} & x & x & x & x \\ \hline
{\it 3} & & & 3.20 & 4.14 & {\it (2,1)} & x & x \\ \hline
{\it 4} & & & & 2.92 & {\it (3,2)} & 4.35 & {\it (2,1)} \\ \hline
{\it 5} & & & & & 2.73 & 3.41 & 3.96 \\ \hline
{\it 6} & & & & & & 2.60 & {\it (4,3)} \\ \hline
{\it 7} & & & & & & & 2.51 \\ \hline
\end{tabular}
\caption{Numerically determined values of the critical mass $m_{**}$. The
symbol `x' indicates that the  corresponding theory has one BPS wall for any
value of the mass parameter $m$ and therefore there is no $m_{**}$.}
\label{table2}
\end{center}
\end{table}
\end{document}